\documentclass[journal]{IEEEtran}
\usepackage[dvips]{graphicx}
\usepackage{url}
\usepackage{subfig}

\graphicspath{{img/}}
\DeclareGraphicsExtensions{.eps}

\begin{document}

\title{Secure Payment System Utilizing\\MANET for Disaster Areas}

\author{Babatunde~Ojetunde,
        Naoki~Shibata,~\IEEEmembership{Member,~IEEE,}
				and~Juntao~Gao
\thanks{Manuscript received March 8, 2017; revised May 28, 2017; accepted September 5, 2017. Date of publication September 28, 2017. This work was supported in part by the Otsuka Toshimi Scholarship Foundation under Grant 16-S38 and Grant 17-S72, and in part by JSPS KAKENHI under Grant JP16K12421, Grant JP15K15981, and Grant JP16K01288. This material was presented in part at the IEEE AINA 2015 conference, South Korea \cite{ref1}. This paper was recommended by Associate Editor D. Akopian. {\it (Corresponding author: Babatunde Ojetunde.)}}
\thanks{The authors are with the Department of Information Science, Nara Institute of Science and Technology, Nara, 630-0192 Japan e-mail: (ojetunde.babatunde.nq3@is.naist.jp; n-sibata@is.naist.jp; jtgao@is.naist.jp).}
\thanks{Digital Object Identifier 10.1109/TSMC.2017.2752203}
}

\markboth{IEEE TRANSACTIONS ON SYSTEMS, MAN, AND CYBERNETICS: SYSTEMS,~Vol.~49, No.12, DECEMBER~2019}{}

\maketitle

\begin{abstract}

Mobile payment system in a disaster area have the potential to provide electronic transactions for people purchasing recovery goods like foodstuffs, clothes, and medicine. Conversely, to enable transactions in a disaster area, current payment systems need communication infrastructures (such as wired networks and cellular networks) which may be ruined during such disasters as large-scale earthquakes and flooding and thus cannot be depended on in a disaster area. In this paper, we introduce a new mobile payment system utilizing infrastructureless MANETs to enable transactions that permit users to shop in disaster areas. Specifically, we introduce an endorsement-based mechanism to provide payment guarantees for a customer-to-merchant transaction and a multilevel endorsement mechanism with a lightweight scheme based on Bloom filter and Merkle tree to reduce communication overheads. Our mobile payment system achieves secure transaction by adopting various schemes such as location-based mutual monitoring scheme and blind signature, while our newly introduce event chain mechanism prevents double spending attacks. As validated by simulations, the proposed mobile payment system is useful in a disaster area, achieving high transaction completion ratio, 65\% - 90\% for all scenario tested, and is storage-efficient for mobile devices with an overall average of 7MB merchant message size.
\end{abstract}

\begin{IEEEkeywords}
Mobile payment system, endorsement, delegation, MANETs, bitcoin
\end{IEEEkeywords}

\IEEEpeerreviewmaketitle

\section{Introduction}

\IEEEPARstart{L}{}arge scale disasters have a major and lasting social and economic impact on people, causing damage that leads to loss of human life, materials and massive economic loss. One of such impact is leaving people in a disaster area without cash-at-hand to purchase necessities like foodstuffs, clothes, and medicine. Although real cash is considered to be the easiest means for carrying out a transaction, it may be impossible to get cash in a disaster situation since access to a bank is restricted both physically (roads may be blocked or the bank destroyed) and electronically (communication infrastructures, like wired networks and cellular networks, may fail due to an earthquake or flooding). Furthermore, existing payment systems require such communication infrastructures for transactions in a disaster area. To enable people to do transactions even in a disaster area, therefore, of vital importance to people in disaster areas is an infrastructureless mobile payment system which can utilize flexible and robust mobile adhoc networks (MANETs) formed via the widely used smart mobile devices (smart phones, etc.).

Furthermore, several payment systems are developed to provide electronic currency services, but none has been specifically created to solve the payment challenges faced by the people in a disaster area (see Section \ref{RelatedWork} for related work). The proposed system is also capable of providing such services, however, since there is no access to the bank in a disaster area, the use of electronic currency for online transaction is restricted. Therefore, our secure payment system is centered on enabling offline transactions utilizing MANETs. In designing such a MANET-based payment system, the following challenges \cite{ref2} should be considered: (1) \textit{Frequent network disconnection -} One of the characteristic of MANET is low-power supply, this can impede a constant connection between users. (2) \textit{Persistent change in topology -} Topology changes quickly in MANET as a result of node's mobility in the network. Thereby leading to a decrease in performance. (3) \textit{Inadequate security -} Secure characteristics of wireless networks are lacking in MANETs; this increases the flaws of MANETs to attacks.

In this paper, we propose a mobile payment system that utilizes self-organized MANETs to enable people to carry out a transaction in disaster areas. The main contributions are summarized as follows.

\begin{itemize}
\item First, we propose a new mobile payment system to allow electronic commerce in disaster areas, in a situation where the bank is not accessible. 
\item Secondly, we introduce an endorsement-based scheme to provide a merchant payment guarantees for a customer using multilevel-endorser scheme to sufficiently cover transaction amount. 
\item We introduce a transaction-log-checking scheme (called event chain) to prevent double spending attack before a transaction is completed. In addition, we propose an electronic money scheme (called e-coin) for account balance checking and to prevent a predetermined number of parties ($N_{c}$) from colluding. 
\item We also adopt a light-weight scheme, based on techniques of Bloom filter and Merkle tree, to reduce communication overheads. 
\item Additionally, we introduce a mutual tracking mechanism that can proof that transaction are valid and reliable. 
\item A digitally signed photo is proposed for authentication and to restrict an attacker from carrying out a fraudulent transaction and impersonating others. 
\item Furthermore, we adopt a blind signature technique to protect user's privacy by ensuring that each user uses different temporary IDs in every transaction. 
\item Finally, we evaluate the performance of our proposed secure payment system by simulation to test the usability in disaster areas. Our simulation focused on: the ratio of successful transaction completions, merchant communication overhead, the validity ratio of event chain, the size of an event chain and the effect of various parameters such as endorser density, mobility speed of nodes and density of monitoring nodes on the transaction completion ratio. Our simulation results showed that the transaction completion ratio increased significantly by an average of 48\%, 28\% and 22\% using 100, 200 and 500 mobile nodes, respectively.
\end{itemize}

The rest of this paper is organized as follows. Section \ref{RelatedWork} introduces related work on mobile payment systems. In Section \ref{prelim} we present preliminaries concerning payment system participants, authentication, user registration, etc. We introduce our proposed secure payment system in Section \ref{overviewofendorsement} and evaluate the proposed system in Section \ref{evaluation}. Section \ref{conclusion} concludes the whole paper.

\section{Related Work}
\label{RelatedWork}

Several studies have been carried out on mobile payment systems which, however, require the support of communication infrastructures to enable secure transactions and are therefore unsuitable for disaster areas without communication infrastructures. Li \textit{et al.} \cite{ref3} introduce an electronic payment mechanism that permits a payment transaction between a vehicle and a merchant when there is a limited connection, however, this mechanism needs a constant link from the merchant to the bank to complete the transaction, and cannot be used, therefore, to provide the needed services for people in a disaster area. Dai \textit{et al.} \cite{ref4} proposed an offline payment mechanism, that is used to buy digital goods. Their proposed mobile payment system adopts mechanisms from Dai's previous works, which introduced a debit-based payment protocol. Patil \textit{et al.} \cite{ref5} introduced an offline electronic coupon micro-payment system. Their scheme is based on credit and allows users to delegate their ability to pay for an item to another person device. The electronic coupon scheme delegation protocol is based on multi-seed payword chains. Their scheme focuses on minimizing the computational cost of mobile devices with limited resources. Similarly, Chen \textit{et al.} \cite{ref6} proposed a scheme that focuses on e-payment systems with electronic cash. To reduce a merchant's burden of having an account for depositing electronic cash received from customers with multiple banks. Chen's scheme introduced the concept of deposit delegation, which allows a merchant to maintain a single account at its trading bank: the system delegates all deposits from various banks into that account. Kiran \textit{et al.} \cite{ref7} introduced a payment system that uses a public-key and a cryptographic hash function to provide security for the transaction. In addition, the proposed payment system uses chains of delegates in which a customer can delegate the authorization to transfer money from the customer's account to other clients (to a vendor, for example). The system allows clients to carry out transactions both on-line and off-line.
 
Hu \textit{et al.} \cite{ref8}, for example, proposed an online micro-payment system where a customer can purchase goods from the merchant. To do this, a customer need to first send to the merchant a purchase request together with the payment authorization. In addition, the identity of users is confirmed indirectly, hence, customer's privacy is protected. However, the protocol can only handle one payment at a time, and relies on a trusted third party, which sometimes hinder the performance of the system. Wang \textit{et al.} \cite{ref9} introduced an electronic cash payment system which reduces the computational overhead of transactions. The computational cost reduction is achieved by integrating the trapdoor hash function into the system. Wang's payment system requires only integer multiplication and addition operations for computation, similar to \cite{ref10,ref11}. 

Chang \textit{et al.} \cite{ref12} focuses on an e-payment system by introducing a novel electronic check scheme to address the inflexibility of the electronic check proposed in \cite{ref13,ref14}. The scheme adopts cryptographic techniques such as a one-way hash function, a blind signature and RSA cryptosystems to protect the system against attacks. The scheme allows a customer to attach the cost of goods to be purchased and the merchant information to the electronic check during a transaction, thereby achieving mutual authentication by the customer and the merchant. Liaw \textit{et al.} \cite{ref15} also adopted a similar concept to Chang's electronic check mechanism to introduce an electronic traveler check scheme that is capable of handling an offline/online transaction. However, Liaw's scheme, unlike Chang's electronic check, adopts a one-way hash function which improves performance and reduces the cost of the system. The customer ID is added to the traveler's check to prevent impersonation of the customer by other users. Dahlberg \textit{et al.} \cite{ref16} survey several existing mobile payment systems and suggests the basis for evaluating the mobile payment study. Furthermore, concerning several gray areas, they propose solutions on which, they suggest, future mobile payment research should be centered.

Nakamoto \cite{ref17} introduced a distributed e-cash system known as Bitcoin that does not depend on a central authority. In the system, a new transaction is transmitted to the entire network, and each node receives the transaction into a block. Then each node attempts to perform a reverse calculation of a hash function as proof-of-work to verify the transaction in their blocks. (The verification procedure is called mining, and each miner are compensated for each block verified). This calculation takes a large amount of computation. Nodes receive a block only if the transactions are genuine and if the Bitcoin has not been used in the previous transaction. The hash of a received block is used in the next block to form a block chain, and with this, all users can agree on the sequence in which transactions occurred. However, Bitcoin requires a device with high power, and transactions are computationally irreversible, so that Bitcoins can never be replaced if a user's private key is forgotten or destroyed. 

This paper differs from related work in the following points: We introduce a secure payment system that utilizes infrastructureless mobile ad-hoc networks (MANETs) to permit users to buy recovery goods in disaster areas. Also, we propose a mechanism that ensures that double spending will be detected before a transaction is completed, unlike existing systems that detect double spending only when e-coins are deposited in a bank or deducted from a customer's account. Our proposed system uses an approach comparable to that of Bitcoin in that transactions are stored in the block chain. However, our method differs in its techniques, since users in our system do not need proof of work. Rather, users calculate the hash value of a transaction log, and neighboring nodes append their signatures to the log to form an event chain (similar to a block chain). The event chain can be verified by surrounding neighboring nodes. Unlike most existing payment systems, our proposed mechanism does not depend on a central authority or mint to detect double spending. 

\section{Preliminaries}
\label{prelim}

In this section, we introduce entities involved in our payment system, user registration and authentication processes, system assumptions and a purchase example in a common payment system.

\subsection{Participants}
\label{part}

All entities (customer, endorser, merchant, and bank) that join and are involved in the payment system will be referred to as users. All users communicate through MANETs. 

\begin{itemize}
\item \textbf{Merchant} - a user that provides goods. 
\item \textbf{Customer} - a user that buys goods from a merchant. 
\item \textbf{Endorser} - a user that agrees in advance to make payments for the customer, if the customer fails to pay. 
\item \textbf{Monitor} - a customer that audits every transaction within the radio range to make sure that each message is valid and reliable.
\item \textbf{Bank} - an organization that maintains users' accounts.
\item \textbf{Delivery Truck} - A truck used for delivering items to the customer. Also used to pass messages between the bank and the users (endorsers) in a disaster area every two days. 
\end{itemize}

\subsection{Proposed System Overview}

Our proposed payment system adopts two operation modes: the first mode is the Internet mode, which functions like every normal online payment system and it is used when there is no disaster. The second mode is the MANET mode, which is used in a disaster situation. When there is a disaster, the system automatically switches the operation mode from the internet mode to MANET mode. Since our goal is to allow people in a disaster area to access essential amenities, we will focus on the disaster mode of our payment system. 

In payment systems, successful transaction completion is essential, however, this cannot be achieved if there is no communication between the users, merchant and the bank. This is the case in a disaster area where the communication infrastructure may be destroyed and access to the bank is cut off both physically and electronically. Therefore, the first aspect of our payment system for the MANET mode is to establish a means of communication among users in a disaster area. To achieve this, we adopt infrastructureless MANETs and DTN- based communication (the communication process is explained later in Section~\ref{sec:comm}.

Then, the next aspect is to establish a means of identifying users and confirming if there is enough account balance to pay for an item since there is no direct connection to the bank during transaction. Therefore, we introduced various schemes such as digitally signed photo and e-coin to achieve this. Also, it is impossible to get physical cash, which a customer needs to pay for the item being purchased since access to the bank is restricted due to destruction of bank infrastructure and communication. Hence, a customer cannot make a direct transaction with the merchant. Detailed implementation of these schemes is explained later.     

\subsection{Registration} 
\label{reg} 

To join the system, customers and merchants register with the bank before a disaster occurs. Each user generates its public and private key pair, then sends only the public key to the bank. The bank is the only trusted party among all the entities involved in the payment system, hence, acts as a certification authority and set the key expiration which can be as long as specified by the bank. Introducing a separate third party to carry out this function will introduce a bottleneck in the system as all users will need to communicate with this third party and since the bank is not available in the disaster area, thereby introducing more overhead in the system. Hence, paying the merchant for a transaction will be difficult. The private key is kept secret by each user. The notations for a user's public and private keys are shown as follows: User's Identity - Bank (${B}$), Merchant (${M}$), Customer (${C}$); Public Key - Bank ($K_{B}$), Merchant ($K_{M}$), Customer ($K_{C}$);  Private Key - Bank ($K_{B}^{ -1}$), Merchant ($K_{M}^{ -1}$), Customer ($K_{C}^{ -1}$) and Digital Signature - Bank ($S_{K_B^{-1}}$), Merchant ($S_{K_M^{-1}}$), Customer ($S_{K_C^{-1}}$).

The registration process in our system can be divided into three stages: merchant registration, customer registration and endorser selection. This registration process takes place before disaster happens.

\subsubsection{Merchant registration}
\label{merchantreg} \

A merchant submits a registration request to the bank to join the mobile payment system. Then the bank accepts the registration request and generates public and private keys for the merchant.

\subsubsection{Customer registration}
\label{customerreg} \

A customer submits a registration request to the bank to participate in the mobile payment system. Then the bank accepts the registration request and generates public and private keys for the customer. The customer selects a photograph and requests the bank to sign the photograph with the bank's digital signature. The bank signs the customer photograph with the bank's digital signature.

\subsubsection{Endorser selection}
\label{endorsersel} \

The customer submits the list of users that will serve as his/her endorsers in the system before disaster occurs (this endorsers are only used for MANET mode transaction). If a user agrees to endorse other specific users, the user deposits real money in the bank. Since there is no direct connection to the bank (both electronically and physically) in a disaster area, the deposited money need to be converted to electronic coins which is used in a disaster area to confirm if an endorser has sufficient money to endorse other user's transaction when the purchase of an item is initiated. The bank generates electronic coins equivalent to the amount deposited by the user (now as endorser).

\subsection{Providing Authentication and Security} 
\label{auth}

In an online payment transaction, the customer identity is verified real-time via the bank, and access to the payment system is allowed providing the verification is successful. A customer cannot be impersonated without an attacker knowing the customer's information. In a disaster area, verifying a customer's identity is currently difficult as a direct link to the bank is not accessible, as a result of the lack of a communication infrastructure.

In our system, each customer chooses a photo that will be digitally signed by the bank, which is used to verify a customer's identity during a transaction and protects the customer when an attacker stole their phone. (Which is similar to checking an individual photo on an ID card, moreover, in our system the merchant will also check the digital signatures of the bank and the customer which is on the photo.). Another form of biometrics authentication mechanism may also be used.

To further secure transactions, each message is digitally signed and encrypted. Thus achieving non-repudiation of transactions. In addition, a monitor can audit every transaction and thus detects an attacker in the network.

\subsection{Assumptions}
\label{sec:intro_assu} 

We make the following assumptions about our mobile payment system.

\begin{itemize}
\item  Fewer than a predetermined number of parties ($N_{c}$) collude to commit fraud. 
\item Users are identified by digitally-signed pictures. 
\item Most of the users do not power off phones very often. This is to discourage users from deliberately switching off their phones in order to carry out an attack. 
\item Most of the phones owned by legitimate users do not share similar location histories, as their Global Positioning System (GPS) coordinates are error bound with a 4.9 - 10 meter range of each other. 
\item Node density is sufficient in most of the locations. 
\item Users can use GPS in almost every location, i.e., we adopt the use of normal GPS for accessing users GPS position since the A-GPS and other positioning technologies used to improve GPS accuracy cannot be accessed due to the destruction of cell towers when disaster happens. 
\item An attacker is not quick enough to get the needed information from the system before the event chain is invalidated (a scheme explained later). 
\item A user can access a bank using the DTN-based communication formed via the delivery truck at least every two days.
\end{itemize}

\subsection{Communication Model}
\label{sec:comm}

In a disaster area, a delay/disruption tolerant network (DTN) \cite{ref18} can be used in addition to a MANET formed among user nodes. The DTN communication can be achieved when two nodes in close proximity to each other communicate. Using the store-carry-and-forward technique, a node stores a message temporarily and forwards the message when the node comes across another node. For the DTN in a disaster area, our approach uses smart phones of users and the delivery truck to form such a network. 

\textbf{Our Network Model:} Since there is no direct communication to the bank as a result of the destruction of the existing communication infrastructure and users in a disaster area are characterized by limited resources (such as bandwidth), it takes several days for users’ messages to get to the bank. We assume that customers and endorsers are in close proximity to a merchant. Therefore, we adopt a network with a communication range of 100m between the users in a disaster area (that is, customers, endorsers and merchant). A minimum of six (6) nodes (i.e. one customer, one endorser, three monitors and a merchant) are required to complete a transaction successfully and the 6 nodes are present within this communication range. When a user sends a message to the merchant, the message is store-and-carry-forward by the intermediate node between the customer and the merchant. In addition to a MANET formed, we introduced DTN-based data dissemination and collection via the delivery truck to transmit messages to/from the bank for the users and the merchant. Each delivery truck moves from the nearby reservoir and cover regions one after the other. The delivery truck is used to deliver items to the merchant in the core disaster area from the nearby reservoir and data moves with this truck. Therefore, with the DTN formed, multi-hop data transfer is possible and communication formed by the truck to the bank in a non-affected area is established. 

\subsection{Payment System in Areas without Disaster}
\label{sec:pay_nodisaster}

In areas that are not affected by a disaster, the customer and merchant can connect directly to the bank using the wired or wireless networks. The steps to purchase an item in such a payment system is illustrated below:

\begin{enumerate}
\item The customer broadcasts a transaction order to purchase an item from the merchant, (for example, an apple that costs \$20). 
\item The merchant verifies the customer's identity and forwards the billing message to the bank, (for example, customer \textit{C} requests to purchase an apple that costs \$20). 
\item The bank confirms the customer's account balance and accepts the transaction if the balance is enough to cover the cost of the transaction. Then withdraw the equivalent cost from the customer's account and inform the merchant to supply the item. However, if the account balance is not enough, the bank rejects the transaction.
\item The merchant delivers the item to the customer.
\item The transaction amount is paid to the merchant, then the bank sends transaction completion notification to the customer.
\end{enumerate}

This approach will be unsuccessful in a disaster area due to the following reasons: 

\begin{enumerate}
\item \textbf{Inaccessible communication infrastructures}. 
\item \textbf{Inaccessibility of a bank}.
\item \textbf{Fraudulent Transactions and Impersonation}. 
\item \textbf{Security/Authentication Issues} | Real-time verification of user's identity is impossible in disaster areas due to the lack of a communication infrastructure. 
\end{enumerate}

To provide a solution to these problems and ensure that there is a payment system that can function in a disaster area, we propose a secure payment system based on endorsement and adopt various mechanisms to secure the proposed system.

\section{Secure Mobile Payment System Based on Endorsement}
\label{overviewofendorsement}

In this section, we first introduce the concept of endorsement and then give detailed explanation of our secure payment system for disaster areas. Our system provides payment guarantees to the merchant in a disaster area where there are no network infrastructures nor direct access to a bank.

\subsection{Endorsement}
\label{endorsement}

In our system, an endorsement is a mechanism by which the endorser agrees in advance to make payments for the customer, if the customer fails to pay. For this, the endorser should have real money deposited in a bank beforehand. An endorser agrees to serve directly as a customer's endorser by signing an endorsement agreement, thereby personally guaranteeing the customer's transaction and pledging to make payment for up to the amount deposited by the endorser for every transaction in which the customer defaults in payment. The endorsement agreement comes with the two conditions that (1) the real money deposited in the endorsement account will be restricted (locked) to endorsing a customer (the locking of the account is effected when the mode of our system is switched from the Internet mode to MANET mode) and (2) the amount endorsed for any transaction has a limit. The endorsement agreement is made during registration prior to a disaster. 

In the proposed method, a minimum of one endorser can successfully endorse a transaction as long as that endorser can cover the payment for the transaction amount. However, to avoid a situation where the endorser is not able to pay for the transaction amount which would lead to a shortage of money to pay the merchant. Therefore, we allow a customer to have multiple endorsers to guarantee each transaction so that the endorsement liability for one transaction is shared among all endorsers. In this way, the risk of endorsing is reduced if a customer purchases an item, but then defaults. To motivate endorsers to cooperate and support the mobile payment system, some part of the transaction amount (e.g., 3\%) is shared among the endorsers as incentives. The percentage of the transaction amount to be used for the incentives is agreed between the bank and the merchant when the merchant joins the mobile payment system. In addition, we introduce multilevel endorsement where an endorser delegates its endorsement capabilities to its own endorser. Each user indicates if they want to participate in such multilevel endorsement at registration, which is before disaster happens. In the multilevel endorsement, when an endorser inherits a transaction from users it normally endorsed, it does so using the exact same endorsement amount agreed to for such user. For example, if user \textit{A} is an endorser to user \textit{B}, and user \textit{B} is an endorser to user \textit{C}. Using the multilevel endorsement when user \textit{A} inherit the user \textit{C's} transaction, for the endorsement to be completed, user \textit{A} needs to sign its signature to show its intention to guarantee the transaction. User \textit{A} uses the actual endorsement amount that is agreed for endorsing user \textit{B} to endorse user \textit{A}. In the multilevel endorsement, each user inheriting a transaction still needs to append its signature on each endorsement. Any transaction without endorsement (i.e. there are no primary endorsers or secondary endorsers) is rejected by the merchant.

\subsubsection{Starting transaction after disaster} \

Once a disaster occurs, a customer and a merchant in close proximity agree to begin a transaction; the users and the merchant meet to establish a connection by exchanging IDs and pictures. The customer sends his/her photograph to the merchant for identification. The merchant compares the photograph with the customer's actual appearance. The merchant also confirms the digital signature of the bank on the photograph. When a customer tries to purchase an item, the exchanged picture is used to identify a customer. The merchant verifies the bank digital signature, timestamp and the customer’s digital signature on the picture. The same procedure is used by all users in the network to identify each other. \

\subsubsection{Transaction Process} \

Through the endorsement mechanism, we realize a mobile payment transaction in a disaster area even when there is no direct access to the bank. For example, let us consider a scenario where an endorser \textit{D} decides to endorse customer \textit{C}. The minimum node density required to complete a transaction is six (6) nodes (one customer, one endorser, three monitors and a merchant). The process for customer \textit{C} for buying an item from a merchant using an endorsement mechanism is illustrated below:
\begin{itemize}
\item \textbf{STEP 1:} Customer \textit{C} broadcasts a transaction order message to purchase an item from the merchant, (for example, an apple that cost \$20). The transaction order message contains a transaction order form, customer \textit{C's} temporary ID, the merchant's ID, the endorser's ID, the bank's ID, the item number, the item quantity, etc. 
\item \textbf{STEP 2:} The merchant checks customer \textit{C's} ID (through a digitally signed photo) and generates a billing message. However, since there is no definite process of confirming customer \textit{C} account balance, the merchant forwards the billing and transaction messages to the endorser, to request that the endorser provides payment security the transaction. 
\item \textbf{STEP 3:} The endorser checks the merchant's ID and customer \textit{C's} ID and generates an endorsement message, signifying that he/she will provide payment security for the transaction by signing the endorsement message with his/her signature. The endorser sends the endorsement message, billing message and transaction order message to the merchant, stating for example, "I agree to provide payment security for customer \textit{C's} transaction of \$20". 
\item \textbf{STEP 4(a):} The merchant checks the endorser's ID and customer C's ID and sends all messages to the bank if the IDs are valid. These messages take two days to reach the bank as there is no direct communication to the bank as a result of the destruction of the existing communication infrastructure. The messages are transmitted from the merchant to the bank using the multi-hop data transfer through the bank truck DTN-based data collection. 
\item \textbf{STEP 4(b):} After sending all the messages to the bank, the merchant immediately supplies the item to customer \textit{C}. The merchant will receive the payment as the transaction is endorsed by endorser \textit{D}. 
\item \textbf{STEP 5(a):} The bank checks the ID of all users and if other information provided is genuine. A few days later, the bank checks the account balance of customer \textit{C} and withdraw the transaction amount (\$20). 
\item \textbf{STEP 5(b):} The equivalent amount of \$20 is paid to the account of the merchant. 
\item \textbf{STEP 5(c):} However, in an instance where customer \textit{C's} account balance is not sufficient to cover the transaction cost, the transaction amount is taken from endorser \textit{D's} account.
\end{itemize}

To prevent unfair - exchange, we adopt transaction settlement and dispute settlement process, where the paid transaction amount is set aside for a particular period during which a customer can report a merchant for not delivering the items purchased. The merchant needs to show proof of item delivery to the customer (usually customer signature collected by the merchant when the customer receives the item). If the merchant fails to do so, the paid amount is refunded to the customer. Hence the merchant is paid if the proof is confirmed to be valid or the dispute period elapse without a customer complaint. The merchant does not need to worry about a customer not paying if the endorsers has guarantee the transaction with valid e-coins.  

Our approach enables electronic commerce in a disaster area despite the restricted communication access to a bank. However, we still encounter the challenges presented in Section~\ref{sec:pay_nodisaster}. We will discuss solutions for each challenge successively.

\subsubsection{Preventing Collusion}
\label{deposit} \

In our mobile payment system, endorsers provide financial security to pay a merchant on behalf of their customers. However, since there is no direct connection to the bank during a transaction, there is a possibility that the endorsers and a customer to collude to cheat in the payment system. In addition, there is a possibility that a customer or the endorsers could draw out money from their accounts before the bank deducts money for the item. Therefore, a method is required to check the endorser account balance before the transaction is completed. We adopt the e-coin technique for the endorsers account balance confirmation.

To be able to purchase an e-coin, a certain amount of money needs to be deposited. The deposited money is locked in an endorsement account, thereby preventing the endorser from using the money to buy an item (that is, the endorser can use the money locked only for endorsement). In a situation where an endorser endorses a transaction and attempts to take away all the account balance from his/her endorsement account before the bank confirms the payment, this attempt will fail as the endorsement account is locked during the disaster mode of our payment system. 

E-coin: The bank generates unique e-coins for an endorser, identical to the tokens in \cite{ref19,ref20} $e_ {T_1} $, $e_ {T_2} $, $e_ {T_3} $,... $e_{T_n}$, for instance, the total amount of the e-coins will be equivalent to the account balance of the endorser. The e-coin contains the endorser's ID, the e-coin ID (signed with the bank digital signature), the e-coin value, and a predefined expiration date.

The reason the expiration date is attached to an e-coin is to avoid the endorser losing money from their account if an e-coin is lost or corrupted while being delivered to the endorser. The bank sets a pre-determined expiration date on the e-coin. The e-coin will be invalid after the predetermined date, if the bank has not received a report from the endorser that the e-coin was received. In the case of invalidity, the bank then issues a new e-coin as a replacement for the lost or corrupted one. If the e-coin is not utilized till it expires, the e-coin turns invalid and cannot be accepted by a merchant. A monitor can prove if an e-coin is still valid or not by confirming the expiration date on the e-coin.

To endorse a transaction, an endorser attaches to an endorsement message, an e-coin equal to the endorsement amount of the transaction. (The e-coin is part of the endorsement message, which is signed by the endorser).

In a situation where the endorsed customer pays for the transaction, the bank will reissue the e-coin to the endorser. Otherwise, the corresponding amount will be deducted from the endorser's account. Thereby, collusion between the customer and the endorser is impeded by checking if there is an e-coin attached to the endorsement message.

When an endorser requests a new e-coin from the bank, the e-coin is either received directly from the bank or transmitted to the endorser through the users available within the radio range. As a result of some communication disruption between the users and the bank, the e-coin may be lost or corrupted while being transmitted. Therefore, we adopt the use of the DTN-based data dissemination and collection via the delivery truck for delivering e-coins to endorsers in a disaster area. The bank delivers new e-coins to endorsers every 2 days via truck. Additionally, multi-hop communication can be used to deliver e-coins from the truck to endorsers. Apart from delivering e-coins to the endorsers, the e-coin truck is also used to bring back to the bank such users' messages as merchant payment and refund of e-coins to endorsers for non-default transactions.

\textbf{Number of Colluding Parties}: In our system, there might be four colluding parties: Customer, Endorser, Merchant and Monitor. We analyze different possible types of colluding scenario formed among these parties (e.g. customer and endorsers, customer and monitor, customer and merchant, endorsers and monitor etc.) in our system. \\
\begin{itemize}
\item \textbf{Two Customers Colluding :} A customer acts as if he/she is an endorser to the other customer. 
\item \textbf{Two Endorsers Colluding :} Such colluding can only happen when one endorser falsely acts as a customer (e.g., has the means of forging the customer credentials) while the other endorser guarantees the transaction.  
\item \textbf{Two Monitors Colluding:} This is the same as endorsers colluding; colluding between two monitors can only happen one falsely disguise as a customer while the other disguise as an endorser. 
\item \textbf{Two Merchants Colluding :} This is conceivable only when the merchants get access to the customer and the endorser credentials (e.g., the customer and the endorser private keys, real IDs, etc.). 
\item \textbf{Customer, and Merchant Colluding :} The goal of this type of colluding is to defraud the endorser. In this colluding, a customer pretends to buy an item, then return the item to the merchant and share the money with the merchant. This form of colluding is difficult to detect as the endorser has genuinely agreed to endorse such transaction. 
\item \textbf{Customer, Endorser and Monitor Colluding :} This type of colluding is possible if the endorser is able to forge an e-coin or reused e-coins already used in previous transactions to defraud the merchant. Also, for this to work the colluding parties needs to have three unique monitor for the endorser. The merchant confirms if the e-coin is been double spent and if the event chain is not broken. Transaction is only allowed if valid e-coins and event chain are used. 
\item \textbf{Merchant and Monitor Colluding :} Similar to other colluding with the monitor, collusion with the monitor by a merchant is hard if the monitor of the transaction is not known beforehand. Hence, this can only happen when there is a limited number of users in the system. Our proposed mechanism dynamically assigns a monitor to check if a transaction is valid before appending their signature on the transaction. 
\item \textbf{Colluding with the Merchant :} Collusion between endorsers and a merchant is not possible if there is no customer. Moreover, it is not possible to forge a customer's digital signature, which is needed for every transaction.
\end{itemize}

\subsubsection{Preventing Double Spending} 
\label{double} \

An endorser may try to spend the same e-coin twice for two different transactions, thereby double spending the e-coin, (i.e., using e-currency twice to pay the same or different people). To prevent double spending in the system and also to ensure that the e-coin is secure, a merchant should be able to check the log for all events in the past associated with the endorser. To do this, the endorser requests other monitoring nodes to sign (with their digital signature) his/her transaction log each time a new event occurs. This will, however, require a lot of communication overhead, since the monitoring node will need to go through the endorser's entire transaction log before signing. Therefore, we propose an event chain with a light weight scheme as a solution to double spending.

\begin{figure}[!t]
\centering
\includegraphics[width=3.5in]{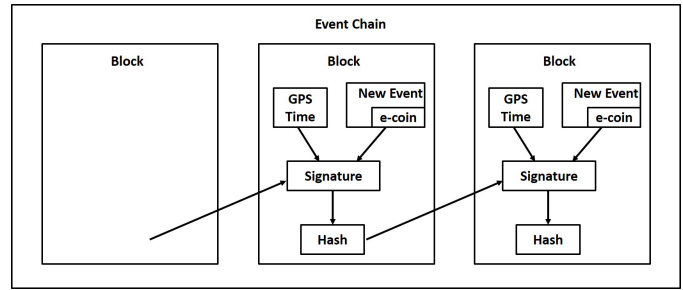}
\caption{Event Chain.}
\label{eventchain}
\end{figure}  

\textbf{Event Chain:} An event chain is a successive application of a cryptographic hash function on a piece of an event log (called a block). Instead of sending and signing on the entire log, the endorser calculates the hash value of the last block, and sends it to a monitor. The monitor signs on the combination of hash value, GPS coordinates, timestamp, and a new event (e.g., spending an e-coin); the monitor then sends the block back to the endorser. In this way, all past events of the endorser are recorded to form an event chain (see Figure \ref{eventchain}), which can be verified by any user. An endorser exchanges a hello message with neighboring monitor nodes periodically to add a new event to the event chain. In order to prevent colluding of up to $N_{c}$ nodes, we require 3$N_{c} + 1$ unique monitor nodes to do this operation since the maximum parties that can collude at a time is three and also, we need to prevent users that are serving as an endorser to a customer from acting as a monitor of the same transaction they are endorsing. Hence the 3$N_{c} + 1$ unique monitor nodes will reduce the likelihood of a monitor node from being compromised as other monitors can verify the same event chain. Using less than 3$N_{c}$ monitor nodes may result into the problem identified in \cite{ref21}, where the two monitor nodes may give conflicting information back to the merchant (i.e., one monitor node validates the event chain while the other invalidates the same event chain). If a predetermined length of time passes after the last event and before a new event is added to the event chain, the event chain is invalidated and can no longer be used. In order to ensure that the e-coin has not been double-spent, a user receives and checks the event log which is the entire event chain from the point at which the e-coin was issued by the bank. When a new e-coin is relayed through multi-hop communication to an endorser, a relay node could possibly duplicate the e-coin before sending it to the endorser. By recording all IDs of e-coins in the event chain, we can prevent the use of a duplicated e-coin. 

Each user keeps the event chain as their transaction log. When a new event is created, a new block is linked to the previous event chain, as shown in Figure \ref{eventchain}. The previous block and the entire log of the present transaction event are signed and forwarded to the monitor. To validate other information in a block, a user requests the entire log. It is possible a user may decide to switch off his/her phone deliberately in order to carry out a reset and recovery attack or to break an event chain. Here the user backup his/her phone, reset the phone to default settings and restore all previous data to buy an item. 

When a phone owned by an endorser is switched off, the event chain is broken as the endorser is not able to exchange hello messages with neighboring monitor nodes. Thereby preventing a new event from being added to the event chain. As a result, the endorser cannot endorse a transaction immediately after turning the phone on but since we assume that there are many endorsers available, the transaction can be guaranteed by other endorsers. The reason we use e-coins only for endorsement is to allow customers to make new transactions immediately after turning off and on the phone since the transaction is guaranteed by his/her endorsers. An endorser on the other hand, first need to exchange hello messages with neighboring monitoring nodes that will verify and update its event chain before such endorser can endorse a new transaction.

By introducing the event chain we can prevent double spending during transaction. However, due to the limited bandwidth of mobile devices in a disaster area, we need to make our mechanism significantly light weight. To achieve this, we adopt the bloom filter mechanism.

\subsection{Light Weight Scheme} 
\label{lightweight} 
We adopt a Bloom filter \cite{ref22} to represent all the spent e-coins since the beginning time of the event chain. That is to say, all spent e-coins are mapped into the Bloom filter. Instead of recording all the IDs of the spent e-coins in the event chain, only the hash value of the latest Bloom filter is recorded in the event chain. When a user checks whether a certain e-coin is double spent, the user receives and checks the Bloom filter.

In the case of a false positive of the Bloom filter, the corresponding e-coin is regarded as already spent; this coin cannot be used. In this case, users have to wait until the e-coin expires and is reissued by the bank. The Bloom filter can represent a set of a sufficiently large number of coins with a small amount of data. When 3000 coins are represented in a Bloom filter with a 1\% false positive ratio, the size of the filter is 4kbytes.  

\begin{figure}[!t]
\centering
\includegraphics[width=3.5in]{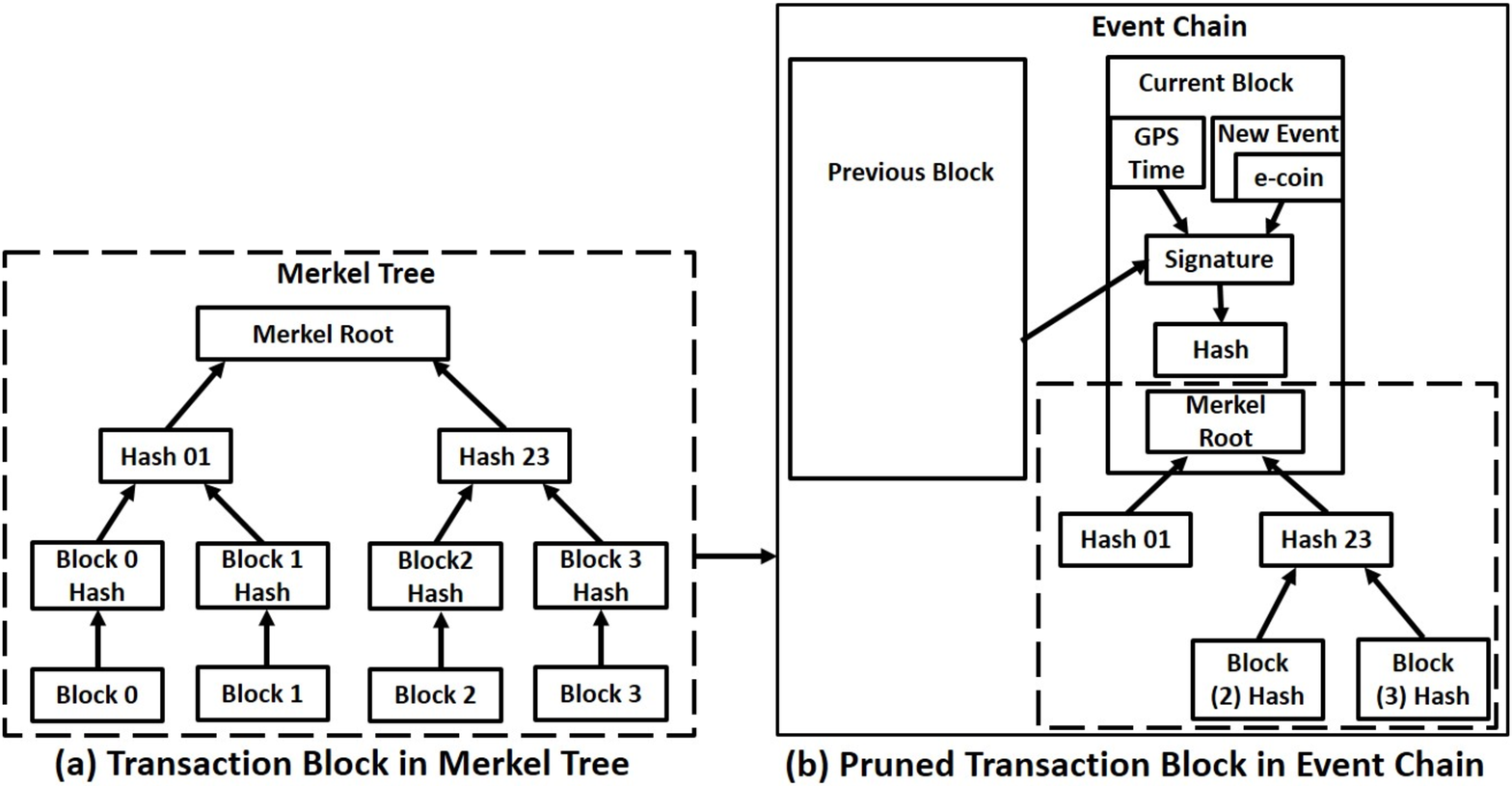}
\caption{Reducing log size using Markel tree.}
\label{eventchainspace}
\end{figure}

We also incorporate the technique called Merkle Tree \cite{ref23} for reducing the size of the transaction block stored in the event chain which is to be checked by the monitor nodes. Each transaction block is hashed and the hash values are then paired together, the resulting paired hash values are further hashed until a Merkle tree root is formed (see Figure \ref{eventchainspace}). The Merkle tree root is stored in the event chain, thereby reducing the size of the event chain. During a transaction, only the reduced event chain and the Bloom filter need to be checked by the monitor nodes.

\subsection{Other Schemes for Secure Transaction} 
\label{schemes}

In this subsection, we explain briefly various schemes adopted to secure transactions in our endorsement-based mobile payment system.

\subsubsection{Location Information-Based Monitoring}
\label{location} \

Many phones might be stolen by one party to use those phones at the same time to attack the system. To prevent collusion using stolen phones, we propose a location information-based monitoring scheme to achieve confirmation of transaction location. According to this scheme, each endorser will continuously exchange HELLO messages with monitoring nodes to prove that the endorser is in a specific location at a specific time. Other users of the system can audit the endorser's transaction location (its coordinates obtained from the GPS of the endorser's phone) by checking the endorser's log of the event chain or the log from the time when the e-coin was received. If an endorser fails to exchange HELLO messages with other users over several time intervals, this would show that the endorser is no more in close proximity of the other users or there is loss of connection. Phones that have similar location histories cannot be used as monitoring nodes.

In addition, if an attacker wants to use a stolen phone, the attacker first needs to find a way to access the customer or endorser's phone which may be protected by a biometric security. Then the attacker will need to break the 1024 encryption key, then get the bank digital signature to forge a new digitally signed picture and the customer digital signatures. 

\subsubsection{Blind Signature}
\label{blindsign} \

Monitoring nodes might access another user's message before signing it during a transaction, thereby compromising the user's anonymity in the system. To prevent this and, more widely, as part of the scheme for preventing a user (customer or endorser) from carrying out multiple transactions using already endorsed transaction order message for reset and recovery attacks, we utilize the techniques of the event chain (to prevent users from reusing the same message) and techniques of the blind signature \cite{ref24} (to protect anonymity).

\subsection{Chains of Endorsers}
\label{chains} 

It is possible that the number of endorsers accessible is not sufficient to pay for the transaction amount, or the customer does not have enough users to serve as his/her endorser, which would lead to a shortage of money to pay the merchant. To detect if there is a shortage of money, the e-coin attached to each endorsement message is checked, this however, would cause the merchant to reject an endorsement message every time the e-coin is less than the transaction amount. To prevent such and to ensure that the customer can purchase an item, even when all the endorsers are not fully accessible or when the endorsement amount are not sufficient to cover the transaction amount, we introduce multilevel endorsement, where each customer has multiple levels of endorsers. When an endorser is not available to endorse a transaction, an endorser of the endorser will be able to endorse such transaction.

According to this method, endorsers have their own endorsers that can inherit transactions to be endorsed. The primary endorser delegates its endorsement capabilities to the secondary endorsers— the secondary endorsers agree to serve as a secondary endorser to a customer beforehand (i.e., during registration, which before disaster happens) and only pays if the primary endorsers do not have enough money to endorse the transaction. The secondary endorsers thereby serve as a proxy to the primary endorsers and are responsible for paying the merchant in a situation where the customer fails to pay for a transaction. The merchant can access the list of the primary and secondary endorsers from the transaction message sent by the customer. The list is created beforehand (during registration) to form an endorsement-chains tree and signed with the bank signature to avoid forgery. The list is updated when primary and secondary endorsers select their own endorsers.

Let's consider a default scenario in which customer \textit{C} buys an item for \$40 from merchant \textit{M}, with endorser \textit{$P_{D}$} as the primary endorser and endorser \textit{$S_{D}$} as the secondary endorser for customer \textit{C}. 

\textbf{Default Scenario 1 :}
When a customer defaults, the primary endorser is billed by the bank. The e-coins are collected from the primary endorser \textit{$P_{D}$} and \textit{$S_{D}$}, however, the secondary endorser is only billed if the primary endorser does not have enough money.

\textbf{Default Scenario 2 :} In a situation when a customer defaults and the primary endorsers do not have sufficient money to cover the payment or are not available during transaction, the secondary endorsers will be charged for the transaction. Let us consider the same scenario described above, the primary endorsers (direct endorser to customer \textit{$C$}) \textit{$P_{D_1}$}, \textit{$P_{D_2}$} and \textit{$P_{D_3}$} do not have enough money. In this case, the secondary endorsers (for example, \textit{$S_{D}$}) are charged. Each secondary endorser is charged according to the amount they agreed to pay for the endorsement.

The merchant sends the billing message to both the primary and the secondary endorsers to obtain their signature on the transaction as a payment guarantee. Unlike our previous method where the merchant searches for the secondary endorsers one level after the other if the primary endorsers are not available, this approach allows the merchant to send the billing message to the secondary endorsers whether the primary endorser is available or not, thereby avoiding the excessive communication needed to search for secondary endorsers when there are insufficient endorsers to endorse a transaction. This way merchant overhead is reduced. This will also ensure that there are more endorsers available to endorse a transaction. In a situation where customer \textit{C} fails to pay for the item purchased, both the primary endorser and the secondary endorser will pay instead.

\section{Evaluation}
\label{evaluation}

\subsection{Security Analysis of the Endorsement-Based Mobile Payment System} 
\label{security}

The attacks considered were selected as likely given the limitations of a disaster area plus other common MANET security challenges. Other MANET-related attacks will be considered in future work.

\begin{itemize}
\item \textbf{Impersonation Attack:} To prevent impersonation, customer \textit{C} attaches a photograph that is digitally signed by the bank before the bank encrypts the message. An attacker cannot impersonate Customer \textit{C} without obtaining his/her digital signature. 
\item \textbf{Colluding Attack:} In a situation where an endorser and a customer collude to cheat in the payment system (e.g., a dishonest endorser may endorse a dishonest customer while neither has money in their accounts). The e-coin technique is used to confirm the endorsers' account balance during the transaction. Endorsers attach to an endorsement message, an e-coin equal to the endorsed amount of that transaction. 
\item \textbf{Double Spending:} Suppose endorser \textit{D} endorsed customer \textit{C} for a transaction with merchant \textit{$M_1$} with an e-coin (for example, $e_ {T_3}$) and then tried to use the same e-coin to endorse another customer's transaction with \textit{$M_2$}. The monitoring user first checks to see if the event chain is broken or valid. If valid, then the monitoring user can hash and sign the event chain. So the event chain prevents an endorser from double spending an e-coin in our system. 
\item \textbf{Non-Repudiation of Transaction Location Source:} Suppose many phones are stolen by an attacker, collusion among those phones is possible. Also, a customer or an endorser current transaction may be carried out from a different geographical location which differs from the location of previous transactions, and then repudiate having made such a transaction. Regarding such cases, other users of the system can detect if any transaction has been carried out away from an endorser's usual location by monitoring the transaction location. The usual location is the geographic location where the user’s phone has been used for a few days. The endorser's entire log of the event chain or the log since an e-coin was received is compared to the event chain at the end of the previous HELLO message exchanged by the endorser. This makes it impossible for an attacker, a customer or an endorser to carry out a transaction in a location other than the usual location. 
\item \textbf{Reset and Recovery Attack:} Suppose a dishonest customer buys an item from a merchant, then resets the phone to the default settings. Then the customer recovers the backup data and uses the same data to buy an item from a different merchant. To reuse a message (a transaction order message or an endorsement message) or an already endorsed transaction message, the user needs to change the event chain of all previous transactions in order to modify the hash values, GPS coordinates and the timestamp in the previous transaction. The user cannot modify the previous transaction message without changing the hash values. The merchant or the monitor will detect that the message has already been used. They do this by checking the entire event chain to see if the predefined time has passed before a new event was added to the event chain. 
\end{itemize}

\subsection{Simulation Configuration}
\label{simconfig}

\begin{figure}[!t]
\centering
\includegraphics[width=2.7in]{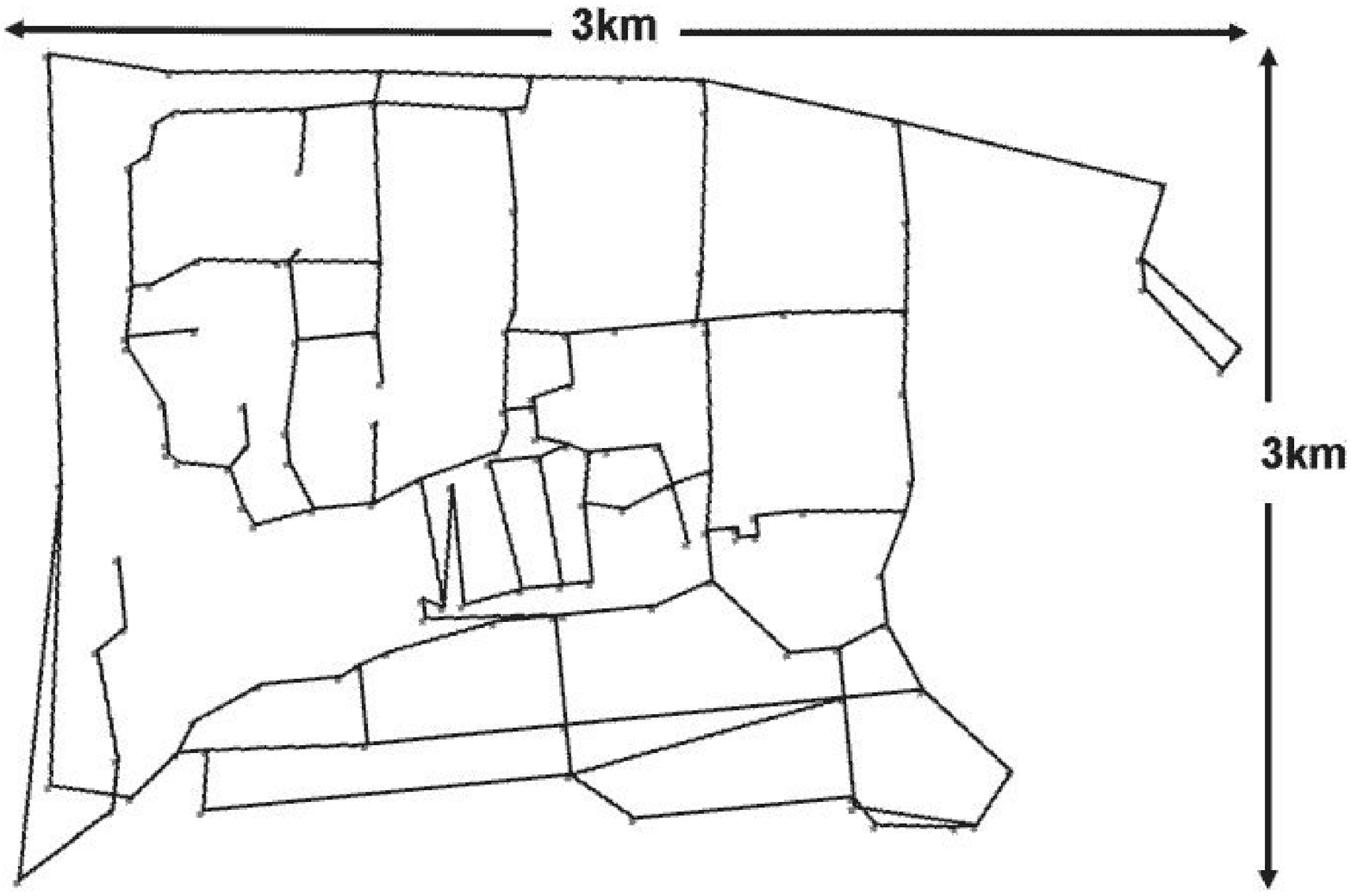}
\caption{Map for Simulation.}
\label{simulationenviron}
\end{figure}

The main objectives of our simulation are to validate (i) usability of our proposed system in a disaster area, and (ii) reduction of communication cost in order to provide excellent service for people in a disaster area.

We conducted our simulation using a customized simulator. The simulated scenario is implemented to enable nodes to connect with each other easily within the transmission range, given that mutual and location monitoring is an important mechanism in our protocol. Mobile nodes are first evenly placed in a 3km x 3km area. This is based on an actual map of the area around Nara Institute of Science and Technology in Nara, Japan, as shown in Fig \ref{simulationenviron}. The skeleton map represents the road network there. Each node moves according to the Random waypoint mobility model \cite{ref25} at a uniform speed of 1 to 1.4m/s and a pause time of 10 seconds. The route is based on Dijkstra's shortest path algorithm. In our simulation, transaction message broadcast time interval was set wherein during this time, the nodes move according to its mobility model and actively perform an action depending on their role at that particular time, (that is, either customer, endorser or monitor). All nodes have the same buffer size and transmission range. We assume 802.11g wireless WiFi (802.11g comes with ad-hoc mode) is used for communication. The summary of the default values used in our simulation is shown in Table \ref{tab:simtable}. The network bandwidth of our simulation is set to 1 Mbps, and our message size is set to 5KB. 

\begin{table}[!t]
\caption{Typical simulation parameters value in a disaster area}
\label{tab:simtable}
\centering
\begin{tabular}{@{}ll@{}}
\hline
\textbf{Parameter}                         & \textbf{Value}  \\ 
\hline
Propagation Model													 & Unit-disc model \\
Bandwidth                                  & 1 Mbps \\
Buffer Size                                & 100-500KB \\
Transmission range                         & 100m \\
Disaster area map size                     & 3km x 3km \\
Number of mobile nodes                     & 100--500 \\
Node speed                                 & 1 - 1.4m/s \\
Node pause time						                 & 10s \\	
Mobility Model                             & Random Waypoint          \\
Message size                               & 5 KB \\
Hello Message Size                         & 5 bytes \\
Hello message Interval                     & 10s  \\ 
Bloom filter size                          & 256bits \\
Proportion of endorser to customer         & 4\%  \\
Number of monitoring nodes                 & 3 \\
Transaction amount (\$)                    & 2 \\	
Endorsement amount (\$)                    & 2 \\
Total e-coin per endorser (\$)             & 3000 \\
\hline
\end{tabular}
\end{table}

The following metrics will be measured in our simulation. 

\subsection{Transaction Completion Ratio} 

\begin{itemize}
\item \textbf{Transaction Completion Ratio (TCR)}: The transaction completion ratio is defined as follows: 
\end{itemize}

\[
   TCR = \frac{\mbox{No. of successful transactions}} {\mbox{No. of transaction messages Rec'd by merchant}} 
\] 

We evaluated the transaction completion ratio to determine the usability of our system in a disaster area. Specifically, we considered two scenarios, the first being the single-level endorsement where transactions are endorsed by primary endorsers only. The second scenario considered is the multilevel endorsement where transactions are endorsed by primary and secondary endorsers. All simulated results in the figures below are the averages from 20 simulation runs (see \cite{ref26}).

\subsubsection{Transaction Completion Ratio of Single-level Endorsement} 
\begin{figure*}[!t]
\centering
\subfloat[Transaction completion ratio]{\includegraphics[width=2.35in]{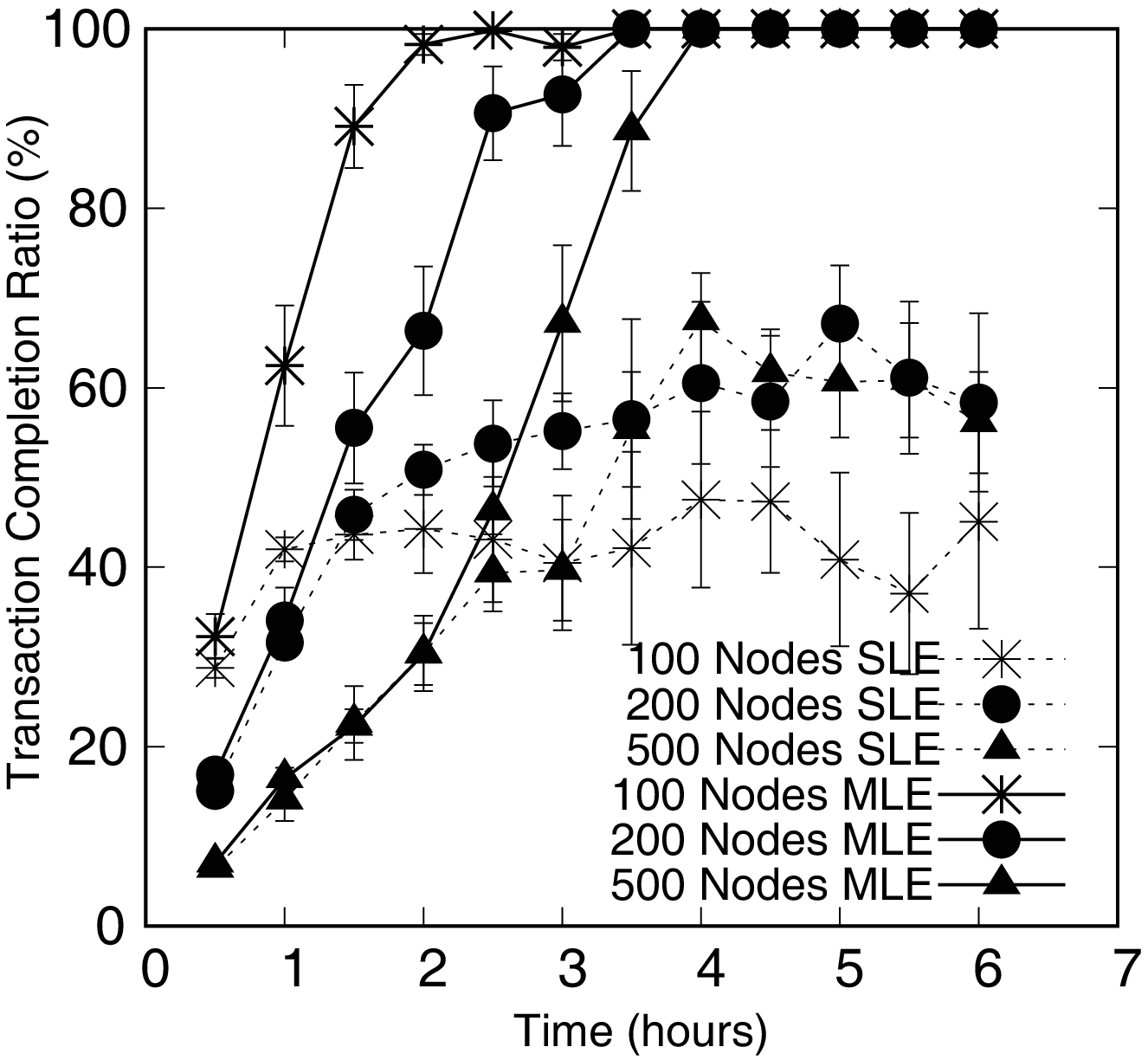}%
\label{tcr}}
\hfil
\subfloat[Merchant message size]{\includegraphics[width=2.35in]{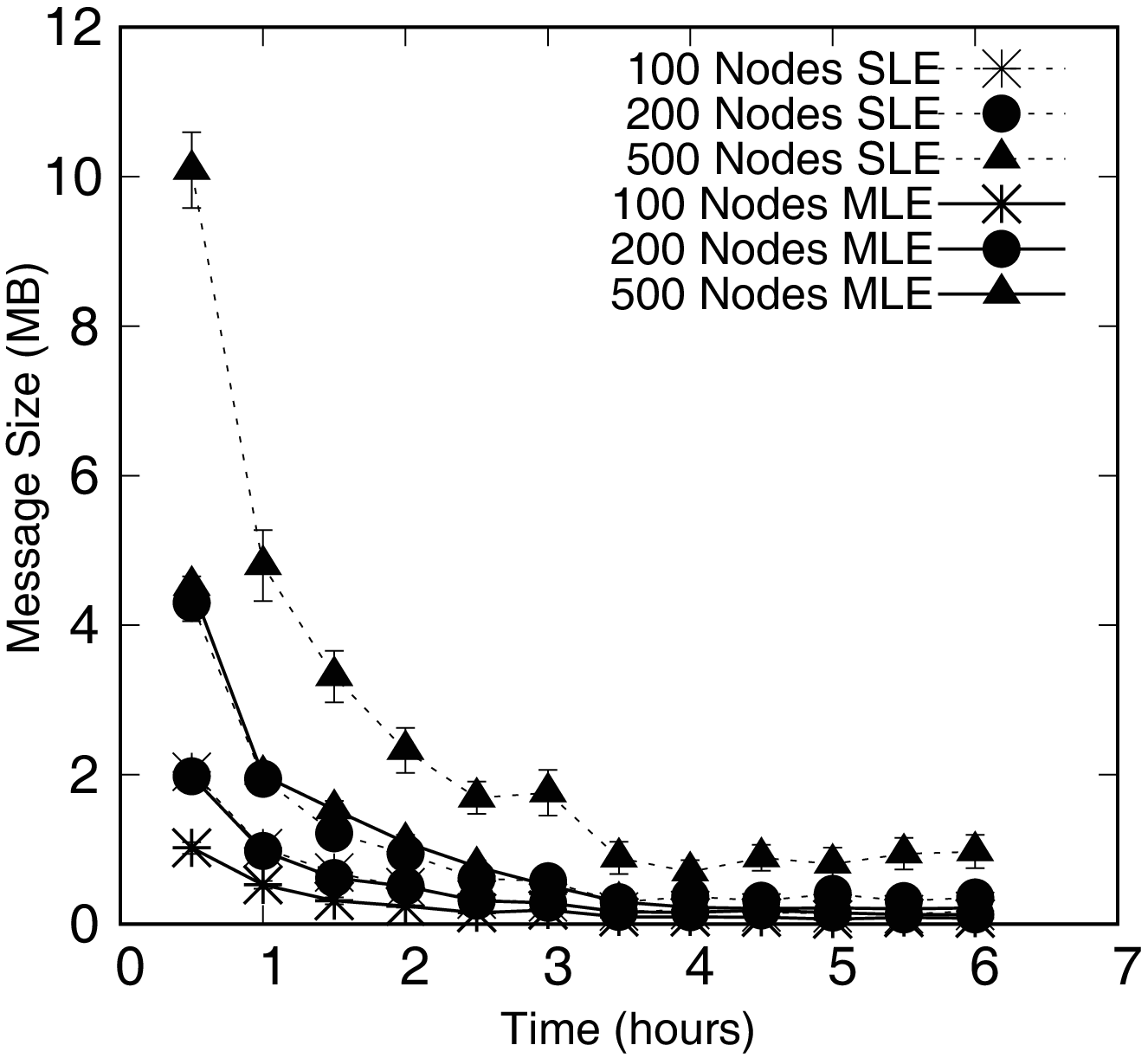}%
\label{merchantoverhead}}
\hfil
\subfloat[Event chain validity]{\includegraphics[width=2.35in]{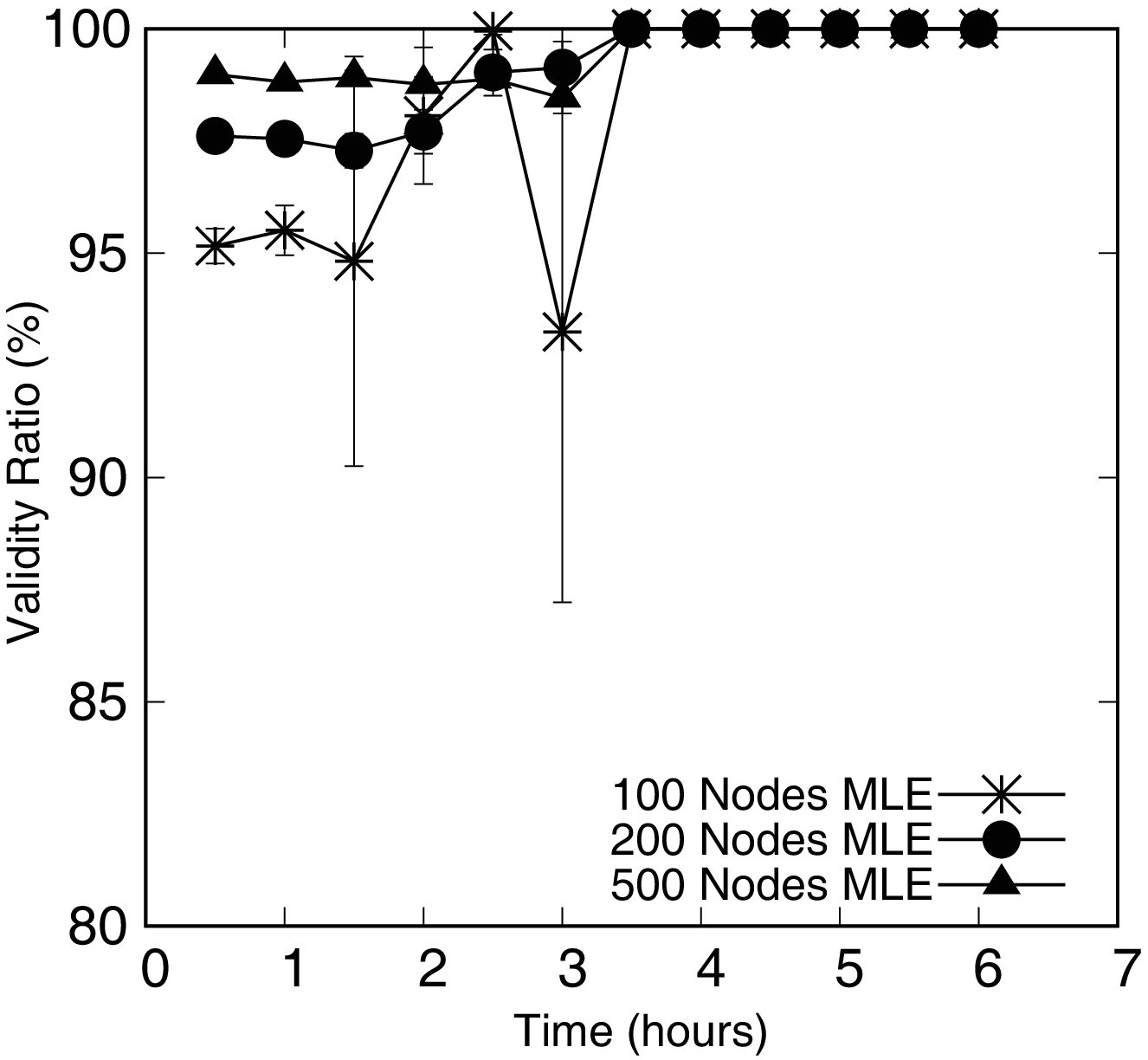}%
\label{fob}}
\caption{Simulation results for the network (SLE : Single-Level Endorsement, MLE : Multilevel Endorsement, endorser ratio = 4\%, merchant No.= 1 and monitoring node No. = 3).}
\label{fig_sim}
\end{figure*}

\
As shown in Figure \ref{tcr}, the single-level endorsement achieved an average of 42\%, 51\% and 43\% of transaction completion ratios for 100, 200 and 500 nodes respectively. The transaction completion ratio increases as time increases at the early stage of the simulation and decreases as the simulation reach a steady stage. This is due to limited number of endorsers for guaranteeing customers transactions.

\subsubsection{Transaction Completion Ratio of Multilevel Endorsement} \
The transaction completion ratio increases significantly with the multilevel endorsement mechanism, averaging 90\%, 80\% and 65\%, respectively, for the three cases above. The significant increase is due to having more endorsers for guaranteeing customers' transactions. Although the transaction completion ratio decreases as the number of mobile nodes increases, the proposed multilevel endorsement achieves better performance when compared with the single-level endorsement, showing an increase from 22\% to 48\%. The significant increase is as a result of having more endorsers to guarantee customer transactions. We achieved this improved performance with the introduction of the multilevel endorsement. We can also observe that the transaction completion ratio increases as time increases, this is because simulations are in a transient stage from 0.5 hours to 3.5 hours, and beyond 3.5 hours simulations reach a steady stage.

\begin{figure*}[!t]
\centering
\subfloat[Endorser density]{\includegraphics[width=2.35in]{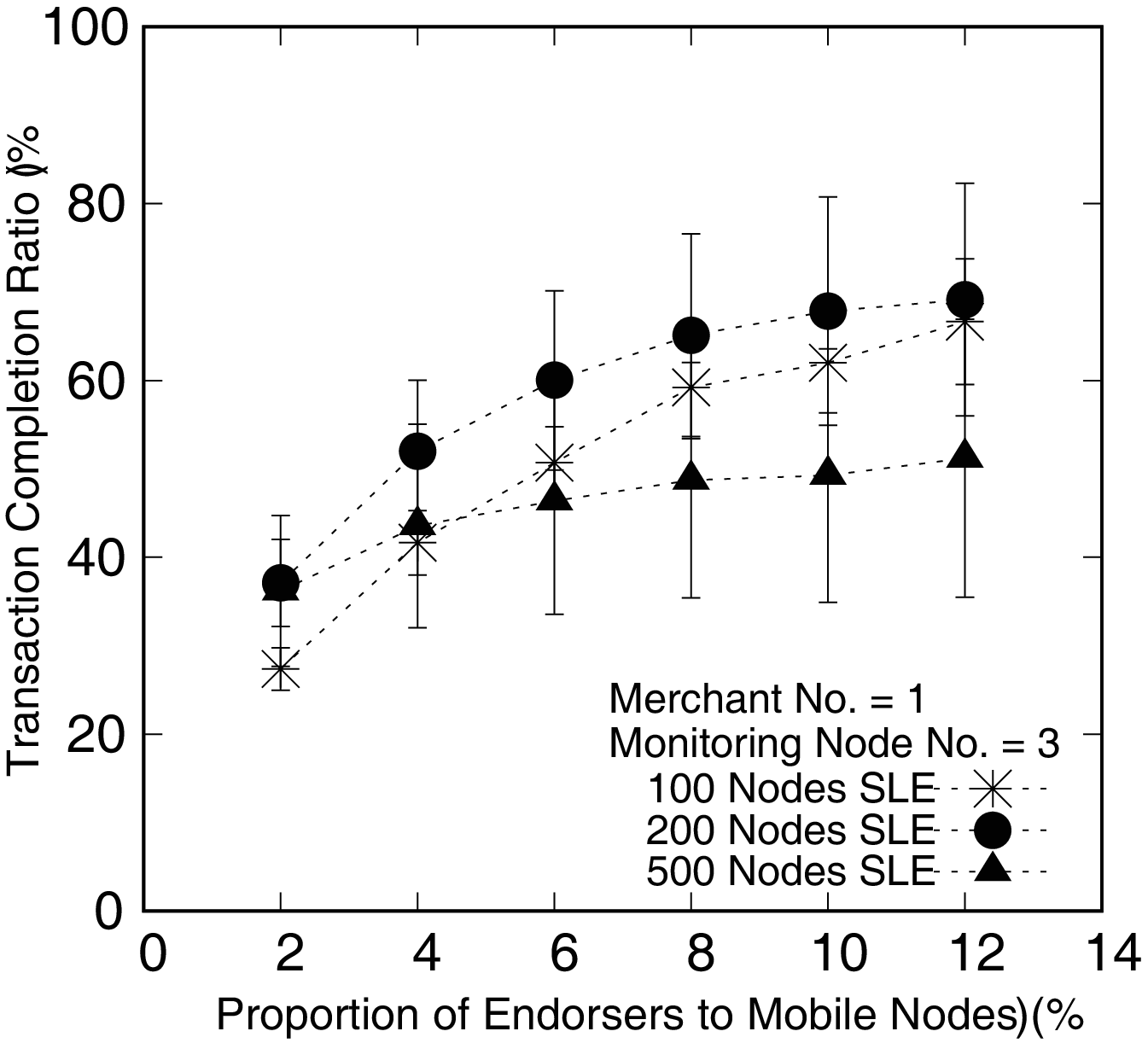}%
\label{endorsedensity}}
\hfil
\subfloat[Mobility speed of nodes]{\includegraphics[width=2.35in]{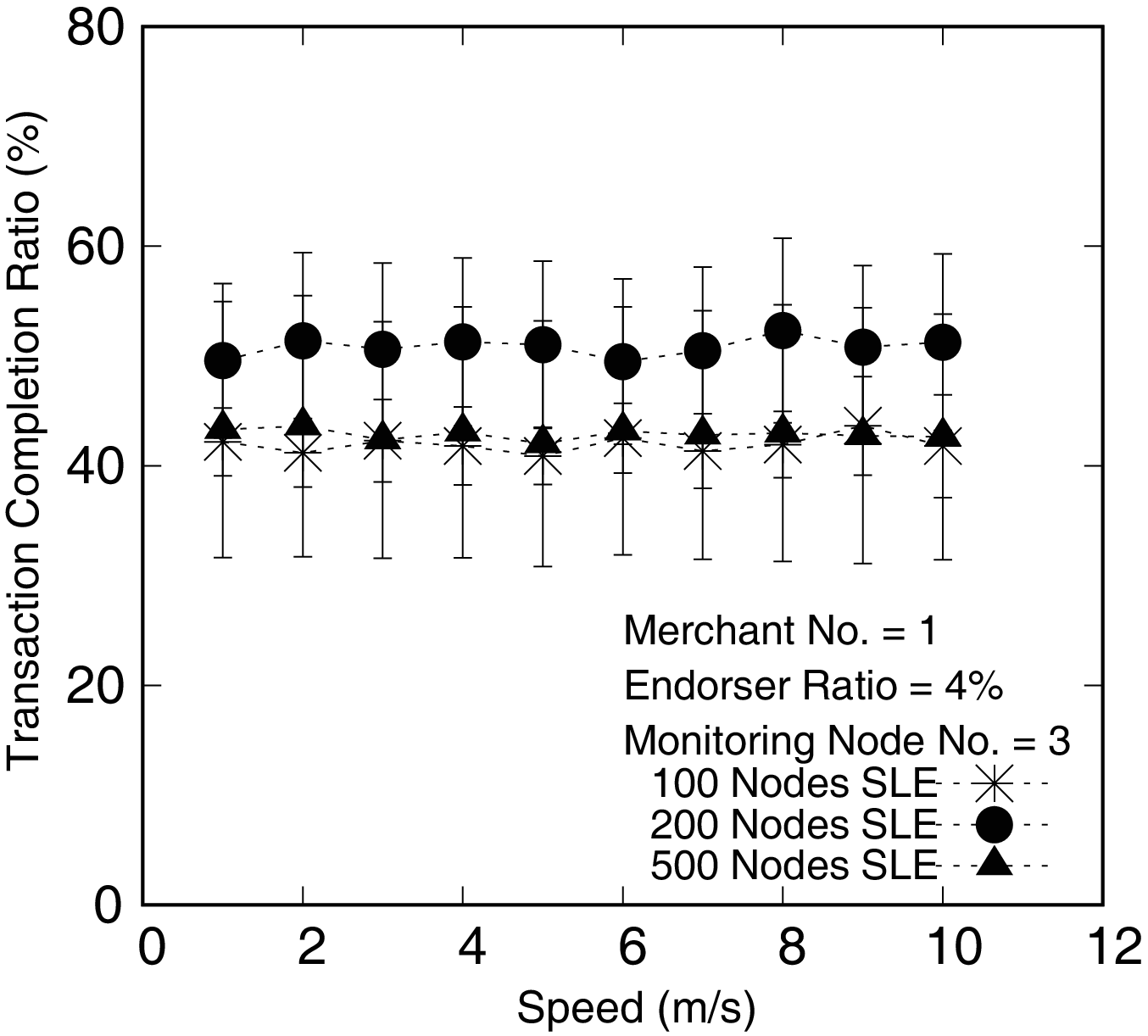}%
\label{nodesmobility}}
\hfil
\subfloat[Density of monitoring nodes]{\includegraphics[width=2.35in]{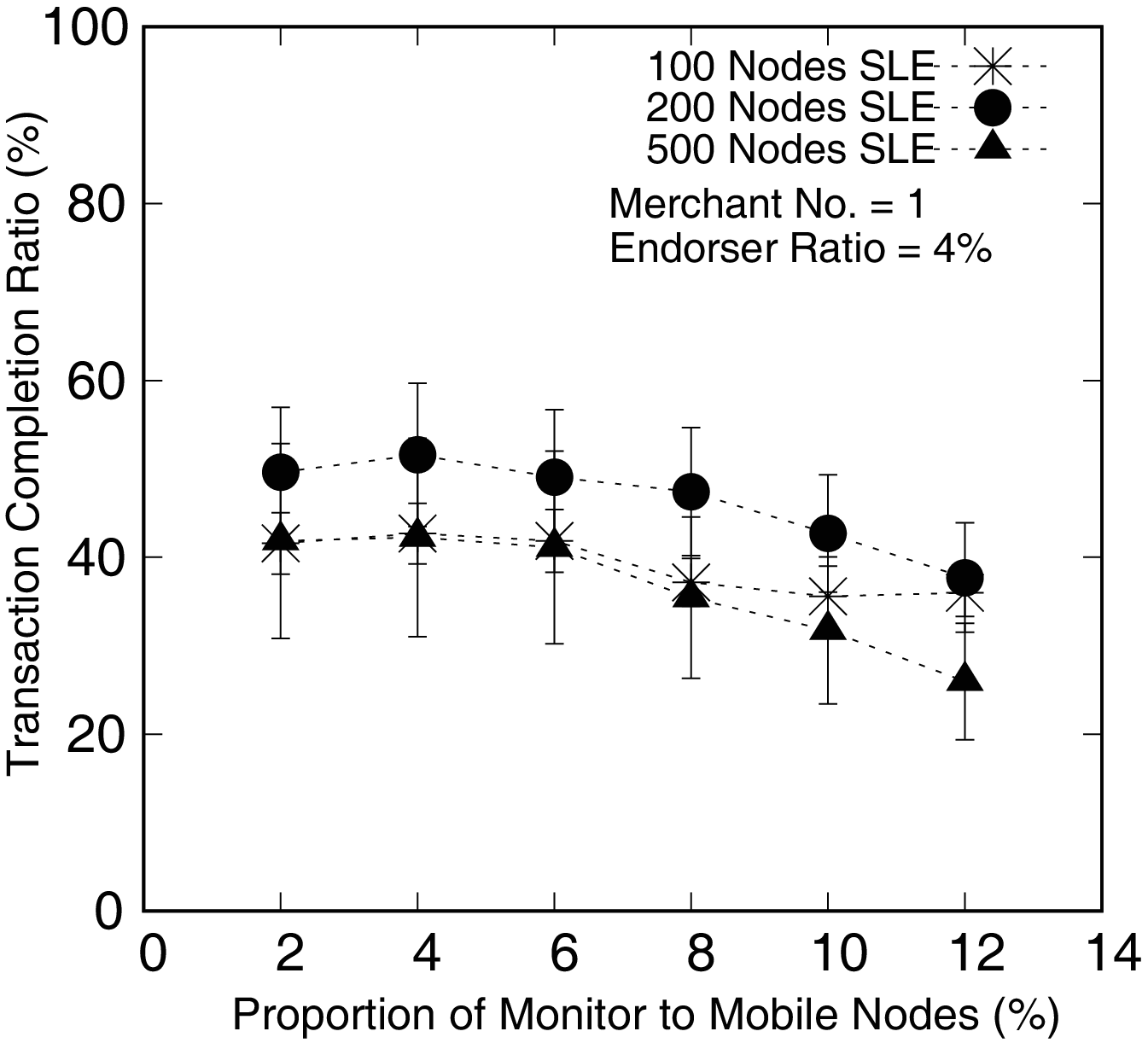}%
\label{monitordensity}}
\caption{Effect of various parameters on transaction completion ratio (SLE : Single-Level Endorsement).}
\label{fig_sim2}
\end{figure*}

\subsection{Communication Overhead}

\begin{itemize}
\item \textbf{Merchant message size}: The size of the message needed by the merchant to check the validity of an event chain and to contact secondary endorsers in a successful transaction. \
\end{itemize}

Our goal in introducing the multilevel endorsement is to increase the transaction completion ratio in our system. However, the merchant should not incur additional communication overhead when multilevel endorsement is used. Therefore, we evaluated the merchant communication overhead of our previous event chain as against the merchant overhead of our proposed multilevel endorsement. As shown in Figure \ref{merchantoverhead}, when compared to the merchant message size in our previous multilevel endorsement where merchant searches for secondary endorsers, one level after the other, there is a 49\%, 52\%, 60\% decrease in merchant message in our mobile payment system with our newly proposed multilevel endorsement for different scenarios with 100, 200 and 500 mobile nodes respectively. In all scenarios, the simulation results show that the overall merchant message size of our system is 7MB on average, with an average of 54\% decrease of that of our previous event chain, indicating that our system with multilevel endorsement is storage-efficient for mobile devices, which have limited resources in disaster areas.

\subsection{Event Chain Validity}

\begin{itemize}
\item \textbf{Validity Ratio of Event Chain (VR):} The ratio at which the event chain is valid in our system, which is computed with the following formula: 
\end{itemize}

\[
\noindent VR = 1 - \frac{\mbox{No. of rejected transaction}} {\mbox{No. of rejected endorsement messages}} 
\] 

Another metric we measured is the validity ratio of event chain. In our mechanism, we introduced event chains to prevent double spending. However, an event chain may be invalidated if dishonest users in the network double-spend e-coins, complete a transaction without e-coins or try to complete a transaction without a monitoring node's signature; or, if too many nodes share a similar location history. The simulation results of the validity ratio of an event chain are shown in Figure \ref{fob}. The results indicate that the validity of an event chain in our system is very high for different scenarios with 100, 200 and 500 mobile nodes with an average validity of 98\%, 99\% and 99\%, respectively, when the proposed multilevel endorsement mechanism is used. The validity ratio increases as the number of mobile nodes increases, which is a result of having more endorsers with valid event chains and having sufficient monitoring nodes available to monitor transactions. The slight decrease observed in event chain validity from 0.5 hours to 3 hours is due to an insufficient number of monitoring nodes when the simulation is in a transient stage.

\subsection{Transaction completion time} 

\begin{itemize}
\item \textbf{Transaction completion time:} The time interval from the time a customer initiates a transaction to the time the merchant accepts the transaction and supplies the items. \ 
\end{itemize}

We also evaluated the transaction completion time of our system in our simulation. First, we explain each process and analyze the simulation time. The customer creates a transaction message, appends its digitally signed picture and its signature to the message. The average computation time for creating the transaction message and generating a signature by a customer is 0.006s. The merchant first verifies the customer information and signature, then verify the digitally signed picture. If the event chain is valid, the merchant generates the billing message and forward it to the endorser, the computational time for this is 0.07s. Similarly the endorser verifies the merchant signature, generate the endorsement message and creates a new block for the event chain with an average computational time of 0.5s. The endorsement message is forwarded to other users for monitoring, a monitoring node validates the event chain and append its signature. The average computational time for monitoring an endorsement message is 0.1s. Finally, the validated endorsement message is forwarded to the merchant and the merchant also validate the endorsement message to avoid collusion between a monitoring node and an endorser, this takes an average of 0.03s. The advantage of our proposed system is that the average transaction completion time is 1.2s, which is the reason for the faster execution of transactions in our endorsement-based mobile payment system.

\subsection{Event chain size} 

\begin{itemize}
\item \textbf{Event Chain Size:} The event chain size with light weight mechanism as against the normal event chain size. \
\end{itemize}

We also evaluate by calculation the size of an event chain scheme when the light-weight mechanism is used. First, we analyze each component that form a block such as new event, timestamp, GPS coordinate, signature and hash value and calculate each components size to get the total size of a block. Each block contains information of 10 e-coins while an event chain using 30 blocks stores information of up to 300 e-coins. Then we calculate the size of event chain. The result shows that the size of the event chain decreased from 3.6KB to 1.7KB with the light weight mechanism, which shows that our light weight mechanism brought a 54\% reduction in event chain size. Similarly, the size of the event chain checked by a monitor decreased from 0.24KB to 0.14KB, with 41\% reduction when the light weight mechanism is applied.

\subsection{Effect of Various Parameters on Transaction Completion Ratio} 

To clarify if our system can achieve better performance than our newly introduced multilevel endorsement mechanism, we examined other scenarios in our simulation by varying different parameters (endorser density, mobility speed of nodes, and density of monitoring nodes) to check how these parameters impact the performance of our system when the single-level endorsement is used.

\subsubsection{Endorser Density} 
Figure \ref{endorsedensity} shows that the endorser's density has an impact on the transaction completion ratio. First, we varied the proportion of endorsers from 2\% to 12\%. The transaction completion ratio increases as the number of endorsers increases, confirming the effectiveness of our multilevel endorsement mechanism. We also observe that there is a slight decrease in the transaction completion ratio for 500 nodes. This decrease is as a result of an insufficient number of monitoring nodes in spite of there being more endorsers in the system, e.g., 40 endorsers, giving an endorser proportion of 8\%.

\subsubsection{Mobility Speed of Nodes} 

Since the contact times of nodes are essential for a successful transaction, we evaluate the impact of a node's mobility speed on the transaction completion ratio. The result is shown in Figure \ref{nodesmobility} with almost constant transaction completion ratios. According to this result, a node's mobility speed has no significant effect on the transaction completion ratio while the mobility speed increases.

\subsubsection{Density of Monitoring Nodes} 

As shown in Figure \ref{monitordensity}, the transaction completion ratio decreases when the number of monitoring nodes needed to complete a transaction successfully increases. The highest transaction completion ratio achieved is found when the monitoring node proportion is set to 4\%. This confirmed the effectiveness of our proposed system setting, i.e., 3 monitoring nodes for validating each message to avoid collusion.

\section{Conclusion}
\label{conclusion}

In this paper, we proposed a new mobile payment system which utilizes infrastructureless mobile ad-hoc networks (MANETs) to enable users to buy recovery goods in a disaster area. According to the endorsement mechanism, endorsers provide absolute payment security for every transaction between a customer and a merchant, therefore permitting mobile transactions in disaster areas even without direct access to the bank. Moreover, by adopting various schemes like the Bloom filter, the blind signature, the event chain, plus location information-based monitoring, the proposed mobile payment system is capable of providing secure transactions, while preventing a fraudulent transaction, collusion, reset and recovery attacks, impersonation of users, double spending. The system also reduces merchant overhead and transaction completion time. 

Simulations confirmed that our endorsement based mobile payment system is useful in disaster areas. Specifically, we evaluated the transaction completion ratio, the merchant communication overhead, the validity ratio of event chain, transaction completion time, and the event-chain size of our system. The multilevel endorsement mechanism in our mobile payment system achieved a better transaction completion ratio, showing an increase of 22\% to 48\% when compared with single-level endorsement. Also, the results show that our system is storage-efficient for mobile devices with limited resources in disaster areas, with an overall average merchant message size of 7MB for all network scenarios tested, which is an average decrease of 54\% compared to our previous mobile payment system. Also, the validity of the event chain mechanism is significantly higher, with an average of 98\% - 99\% for all network scenarios.

\begin{IEEEbiography}{Babatunde Ojetunde} 
received his HND degree in Computer Science from Osun State Polytechnic, Iree, Osun, Nigeria in 2003 and also received his PGD in Computer Science from Lagos State University, Ojo, Lagos, Nigeria in 2006. He received M.E in Information Science from Nara Institute of Science and Technology, Japan in 2015. Currently, he is a Ph.D. student at the Graduate School of Information Science, Nara Institute of Science and Technology, Japan. His research interests include mobile computing, and  ubiquitous computing. He is a member of IPSJ.
\end{IEEEbiography} 
\begin{IEEEbiography}{Naoki Shibata}
is an associate professor at Nara Institute of Science and
Technology. He received the Ph.D. degree in computer science from
Osaka University, Japan, in 2001. He was an assistant professor at
Nara Institute of Science and Technology 2001-2003 and an associate
professor at Shiga University 2004-2012. His research areas include
distributed and parallel systems, and intelligent transportation
systems. He is a member of IPSJ, ACM and IEEE.
\end{IEEEbiography}
\begin{IEEEbiography}{Juntao Gao} 
received his B.S. and M.S. degrees both in Computer Science from Xidian University, Xi’an, China, in 2008 and 2010, respectively, and received his Ph.D. degree from Graduate School of Systems Information Science at Future University Hakodate, Japan, in 2014. He is an assistant professor at Graduate School of Information Science at Nara Institute of Science and Technology, Japan. His research interests are in the areas of performance  modeling  and  analysis,  stochastic optimization and control in wireless networks, queueing theory and its applications. 
\end{IEEEbiography} \vfill

\end{document}